\documentclass[12pt, letterpaper]{article}
\usepackage[letterpaper,top=1in,bottom=1in,left=1in,right=1in]{geometry}

\usepackage[utf8]{inputenc}
\usepackage[english]{babel}
\usepackage{amsmath}
\usepackage{graphicx}
\usepackage{enumitem}   
\usepackage{amssymb}
\usepackage{amsthm}
\usepackage{diagbox}
\usepackage{multirow}
\usepackage{hyperref}
\usepackage{longtable}
\usepackage[normalem]{ulem}
\useunder{\uline}{\ul}{}
\usepackage{lipsum}
\usepackage{mathtools}
\usepackage{tikz}
\usepackage{float}
\usepackage{subcaption}
\usepackage{setspace}
\usepackage{url}
\usepackage{xurl}

\usepackage[style=apa,backend=biber, uniquename=false, ]{biblatex}
\usepackage[T1]{fontenc}
\usepackage[utf8]{inputenc}
\usepackage{array}
\usepackage{verbatim}
\usepackage{url}
\usepackage{amsmath}
\usepackage{graphicx}
\usepackage{geometry}
\usepackage{setspace}

\usepackage{booktabs}
\usepackage{babel}
\usepackage{csquotes}

\usepackage{authblk}

\title{ Human or Robot? Evidence from Last-Mile Delivery Service}

\author[1]{\Large Baorui Li}
\author[2]{Xincheng Ma}
\author[3]{Brian Rongqing Han}
\author[4]{Daizhong Tang}
\author[5]{Lei Fu}

\affil[1]{The University of Texas at Dallas}
\affil[2]{Hong Kong University of Science and Technology}
\affil[3]{University of Illinois Urbana-Champaign}
\affil[4]{Tongji University}
\affil[5]{Alibaba Group Inc.}

\date{}

\begin{document}
\maketitle

\newpage
\begin{abstract}
As platforms increasingly deploy robots alongside human labor in last-mile logistics, little is known about how contextual features like product attributes, environmental conditions, and psychological mechanisms shape consumer preference in real-world settings. To address this gap, this paper conducts an empirical study on consumer choice between human versus robot service, analyzing 241,517 package-level choices from Alibaba's last-mile delivery stations. We identify how product privacy sensitivity, product value, and environmental complexity affect consumer preference. Our findings reveal that consumers are significantly more likely to choose robot delivery for privacy-sensitive packages (11.49\%) and high-value products (0.97\% per 1\% increase in value), but prefer human couriers under adverse weather conditions (1.63\%). These patterns are robust to alternative specifications and controls. These results also underscore that delivery choices are shaped not only by functional considerations but also by psychological concerns, highlighting the need for context-aware service design that aligns strategies with consumer perceptions.

\end{abstract}

\newpage
\section{Introduction} \label{sec:introduction}
The rapid advancement of artificial intelligence and automation technologies has led to increasing integration of service robots into traditional offline service environments. \textcite{Deloitte2023} reports that autonomous delivery, automated and intelligent assistants are rapidly entering industries such as retail, logistics and food service, signaling a new era in which human and robot services coexist. This shift raises a critical yet underexplored question: how do consumers choose between human and robot services? Understanding not only what consumers choose, but also why is essential for firms seeking to optimize service design, resource allocation, and consumer experience in such hybrid human-robot systems. However, the psychological drivers behind consumer preferences and their behavioral consequences have received limited systematic investigation in both academic research and industry practice.

Prior research has primarily examined human-robot interaction in task execution, focusing on the roles of experts, workers, or consumers in conjunction with AI systems \autocite{fugener2021will, wang2023knowledge, lu20241+, zhang2023consumer} as well as human responses to AI-powered robots relative to traditional human or rule-based service providers \autocite{luo2019frontiers, cui2022ai, bai2022impacts, xu2024identity}. However, little is known about how consumers actively choose between human and robot services in real-world contexts, and what contextual or psychological factors shape these choices. This gap is particularly salient in complex offline environments where AI-enabled robots are increasingly deployed, yet empirical research remains limited \autocite{jeon2022last, wang2024beyond}.

Our study addresses this gap by analyzing real-world consumer choices in the context of Alibaba's last-mile logistics network, Cainiao Station. Leveraging a large-scale proprietary dataset comprising 241,517 package-level records in Cainiao campus stations at 32 universities across China, we examine how key contextual features -- product privacy sensitivity, product value and environmental complexity -- systematically influence consumer choices.

Building on prior theoretical research, we propose that different contextual factors activate distinct psychological mechanisms that shape consumer decision-making. Specifically, product privacy sensitivity triggers heightened privacy concerns, leading consumers to prefer less socially intrusive options such as robots \autocite{Smith1996, Awad2006}. In contrast, high-value products can trigger different consumer concerns: some may emphasize the risk of service failure and prefer human couriers for their discretion and adaptability \autocite{Bauer1960, Featherman2003, ma2025understanding}, while others may prioritize consistency and traceability, favoring robots in structured environments \autocite{Wirtz2018, yalcin2022thumbs, you_trusting_2024}. Finally, environmental complexity, such as bad weather, increases the perceived uncertainty, potentially enhancing reliance on human service providers who are viewed as better able to exercise adaptive judgment \autocite{Mayer1995, McKnight2002}.

Based on our analysis of delivery choices, we empirically test our hypotheses and find that user \footnote{The terms \textit{user} and \textit{consumer} are used interchangeably in this paper.} choices systematically respond to contextual triggers. Privacy-sensitive products increase the likelihood of robot delivery by 11.49\% ($p<0.001$), consistent with the notion that consumers seek to avoid social judgment by opting for less intrusive service. For high-value items, a 1\% increase in package value leads to a 0.97\% increase in the likelihood of choosing robot delivery ($p<0.001$), suggesting that robots are perceived as secure and traceable in structured environments. In contrast, adverse weather conditions reduce the likelihood of robot delivery by 1.63\% ($p<0.001$), reflecting users' greater trust in human couriers when facing environmental uncertainty.

Furthermore, we identify the heterogeneity in consumer preferences driven by user characteristics, specifically gender and prior usage experience. Female users are significantly more likely to choose human delivery under adverse weather conditions, with the likelihood of robot delivery decreasing by an additional 1.38\% ($p<0.001$). By taking the interaction between main factors, we find that bad weather conditions attenuate the positive effect of privacy-sensitive packages on robot choice. The likelihood of choosing a robot in adverse weather for these packages was 6.37\%, a significant reduction from the baseline effect of 11.49\% $(p<0.05)$. Prior delivery experience also plays a pivotal role in shaping current choices: selecting robot delivery in the previous orders increases the likelihood of repeating that choice by 24.6\% ($p<0.001$), while prior human delivery reduces it by 49.7\% ($p<0.001$). We also find evidence of cumulative habit formation: each additional prior robot delivery increases the likelihood of choosing robot service by 0.07 -- 0.19\% ($p<0.001$), and 0.16 -- 0.18\% ($p<0.001$) for that of human service. The recalibration effect of cumulative experience is also uncovered by incorporating the interaction terms. These findings highlight the dynamic and adaptive nature of consumer decision-making in hybrid human-robot service systems.

Our study makes several contributions. First, to the best of our knowledge, it is the first to systematically examine actual consumer choice between human and robot delivery service in a real-world offline environment. By utilizing large-scale package-level data, we provide empirical insights into consumer behavior in the emerging phase of human-robot service coexistence. Second, we advance the theoretical understanding of robot service choice by showing how contextual factors activate distinct psychological mechanisms, including privacy concerns, perceived risk, and trust expectations. This framework enriches the literature on human-robot interaction by identifying the conditions under which automation is preferred or resisted. Finally, our findings offer actionable implications for platforms designing hybrid human-robot service systems, emphasizing the need for context-sensitive resource allocation and service channel management in the evolving landscape of automation.

The remainder of the paper is structured as follows. Section \ref{sec:literature review} reviews the related literature; Section \ref{sec:hypothesis development} develops our main hypotheses; we introduce our empirical setting and dataset in Section \ref{sec:empirical setting and data} and present the main result in Section \ref{sec:main results}; Section \ref{heterogeneous analysis} and Section \ref{robustness check} show our heterogeneous analysis and robustness checks respectively; finally, Section \ref{sec:conclusion} concludes.

\section{Literature Review} \label{sec:literature review}
Our work intersects two main research streams: human-robot services interaction and last-mile delivery.

\subsection{Human-robot Services Interaction}
Prior studies have examined both how robot augments the user's work and how users respond to AI-enabled robot services. The former highlights robot's ability to support judgment across domains such as clinical diagnosis \autocite{yan2021ai}, drug development \autocite{lou2021ai}, financial forecasting \autocite{fu2021crowds, lu20241+}, and consumer co-creation \autocite{zhang2023consumer}.

The latter stream of interest focuses on how users evaluate and interact with robots, emphasizing concerns around trust \autocite{dietvorst2015algorithm, xu2024identity}, privacy \autocite{ goldfarb2011privacy, holthower2023robots}, cost \autocite{castelo2023understanding} and empathy \autocite{yalcin2022thumbs}. For example, \textcite{dietvorst2015algorithm} and \textcite{you_trusting_2024} address that trust-in-robots determines user acceptance of robot service directly. \textcite{yalcin2022thumbs} finds that consumers perceive robot to be less empathetic than human ones in emotionally charged scenarios. This agrees with the complementary finding of \textcite{Mende2019} that consumers are more willing to be served by humans when consuming "healthy product" like healthy food due to the experience of high-level social belongingness. However, the belief that "robots do not judge" becomes a positive driver for consumers to conversely choose robot service from the perspective of privacy. \textcite{holthower2023robots} identifies a key situation where consumers prefer service robots over humans if they encounter an embarrassing service, showing that consumers feel less judged by a robot when, for example, purchasing a sensitive product like sexual disease medicine. 

Much of this literature has studied digital or simulated chatbots, recommender systems, or automated work assignment tools, where users passively receive outputs \autocite{luo2019frontiers, cui2022ai, ma2025understanding}.  Relatively little is known about how users make active choices between robot and human service options. \textcite{ma2025understanding} is closely aligned with our research focus, showing that customers in luxury service settings have a lower preference for service robots compared to mainstream settings. This "robot disadvantage" is driven by the customer's desire to feel superior to the service provider, a feeling that human interaction can satisfy but robot interaction cannot. Many of these studies use surveys to understand consumer psychological processes, but have investigated consumer choice in real-choice settings less. We contribute to this gap by analyzing delivery decisions in a physical hybrid human-robot service system, where both robot and human services are available.

\subsection{Last-mile Delivery}

This study also connects to the literature on last-mile delivery contextually by offering empirical insight into how consumers choose between robot and human delivery. Existing work has primarily focused on optimizing delivery network design and efficiency \autocite{fatehi2022crowdsourcing, liu2021time, carlsson2021provably} as well as cost \autocite{qi2018shared, raghavan2024driver}. 

Empirical studies have highlighted the value of operational innovations such as pickup stations and human couriers \autocite{han2024connecting, amorim2024customer, lim2023right, lu2023value}. For instance, \textcite{lu2023value} investigates the value of human delivery in the last-mile delivery context, indicating that last-mile delivery significantly increases sales and customer spending on online retail platform. More recent research has begun to examine robot delivery in field settings. \textcite{han2024value} shows that the introduction of robot delivery increases order volume relative to self-pickup and may generate greater value than human couriers. These studies quantify the aggregate benefits of automation but do not examine consumer-side decisions when both delivery services are simultaneously available. 

Table \ref{tab:positioning} displays how we differentiate our work from prior literature. Using user choice data, we study the decision-making setting directly. Our analysis links contextual drivers to delivery selections, extending the last-mile literature to account for user agency in hybrid human-robot service environments. In doing so, we complement existing supply-side and design-focused work with a demand-side perspective on the adoption and comparative appeal of robot delivery.

\begin{table}[ht!]
\centering
\caption{Positioning of the current study in last-mile delivery literature}
\footnotesize
\label{tab:positioning}
\begin{tabular}{cccc}
\hline
\multicolumn{1}{c}{\begin{tabular}[c]{@{}c@{}}Comparison\\ (row $vs.$ col)\end{tabular}} &
  Self-pickup &
  Human Delivery &
  Robot Delivery \\ \hline
\begin{tabular}[c]{@{}c@{}}Self-pickup\\ (intro. of last-mile sta.)\end{tabular} &
  \begin{tabular}[c]{@{}c@{}}Value of last-mile station\\ \autocite{han2024connecting}\end{tabular} &
  --- &
  --- \\
Human Delivery &
  \begin{tabular}[c]{@{}c@{}}Impact of to-door delivery\\ \autocite{lu2023value}\end{tabular} &
  --- &
  --- \\
Robot Delivery &
  \begin{tabular}[c]{@{}c@{}}Effect of robotic delivery\\ \autocite{wu2023value}\end{tabular} &
  The current study &
  --- \\ \hline
\end{tabular}
\end{table}

\section{Hypothesis Development} \label{sec:hypothesis development}

The increasing integration of service robots into traditional offline environments raises important questions about how consumers make service type choices in human-robot coexistence settings. Based on prior research in service marketing and technology adoption, we propose that consumer choice between human and robot services is systematically influenced by contextual factors that activate different psychological mechanisms. Among these factors, especially we propose that \textit{product privacy sensitivity}, \textit{product value}, and \textit{environmental complexity} act as key contextual triggers that respectively heighten privacy concerns, risk perceptions, and trust expectations. These psychological processes, in turn, shape consumer preferences between human and robot.

\subsection{Product Privacy Sensitivity}

Privacy concerns are key drivers of consumer behavior in technology-mediated services, especially when sensitive products are involved \autocite{Smith1996, Awad2006}. In last-mile delivery, items like intimate apparel or personal care products may evoke anxiety due to anticipated embarrassment \autocite{Argo2005}. Robot service providers offer a distinct advantage in such contexts: unlike human servers, robots are perceived as lacking social judgment capabilities and emotional evaluation, thereby serving as "safe" recipients of sensitive information \autocite{Wirtz2018, Mende2019} and reducing embarrassment when acquiring sensitive products \autocite{pitardi2022service, holthower2023robots}. This perception reduces the psychological discomfort associated with privacy-sensitive transactions and may make consumers more willing to opt for robot service.

\textbf{H1:} 
\textit{For privacy-sensitive products, consumers are more likely to choose robot service over human service.}

\subsection{Product Value}

Consumer's delivery preferences for high-value products can be influenced by distinct theoretical mechanisms. According to classic theories of perceived risk, greater product value heightens the perceived consequences of service failure, thereby increasing the demand for human judgment, discretion, and accountability \autocite{Bauer1960,Featherman2003}. From this perspective, human couriers may be viewed as better equipped to handle valuable items, particularly in the face of unexpected challenges. Prior studies also find a robot disadvantage such that consumers see human staff as key to the prestige of luxury, so robot preference is lower in luxury contexts \autocite{ma2025understanding}. Based on this reasoning, we propose the following hypothesis:

\textbf{H2a:} \textit{For high-value products, consumers are more likely to choose human service over robot service.}

Alternatively, research on service automation suggests that, in highly structured and monitored environments such as campus logistics systems, robots may be perceived as more consistent, traceable, and less prone to error. Once trust in an automated system is established, it can be preferred for important, high-stakes tasks \autocite{you_trusting_2024}. These attributes can enhance user's sense of control and security, especially when the stakes are high \autocite{Wirtz2018,yalcin2022thumbs}. This leads to an alternative hypothesis:

\textbf{H2b:} \textit{For high-value products, consumers are more likely to choose robot service over human service.}

\subsection{Environmental Complexity}

Achieving a high level of autonomy requires adaptation to unpredictable and changing environments, supporting the rationale that humans are currently more capable in such scenarios \autocite{beer2014toward}. In stable settings, consumers may prefer robots for their efficiency and consistency. However, as unpredictability rises -- such as during extreme weather -- trust expectations increase, humans are often viewed as more capable of exercising judgment, empathy, and flexibility \autocite{Wirtz2018}, prompting a shift in preference toward human service providers. 

\textbf{H3:} 
\textit{In complex service environments like extreme weather conditions, consumers are more likely to choose human service over robot service.}

In summary, we propose a framework wherein product privacy sensitivity, product value, and environmental complexity activate distinct psychological processes that systematically influence consumer preferences. Our empirical study in the last-mile delivery offers a unique opportunity to test these hypotheses in a real-world human-robot coexistence environment.

\section{Empirical Setting and Data} \label{sec:empirical setting and data}

By the end of 2024, Alibaba's logistics subsidiary, Cainiao Smart Logistics Network, had deployed thousands of campus-based last-mile distribution facilities, "Cainiao Stations" (see Appendix A Figure \ref{fig:station}), with each university typically hosting one station to enable on-site package collection. Many stations have incorporated last-mile delivery services via both human couriers and robot vehicles (see Appendix A Figure \ref{fig:robot}), where users can choose preferred delivery services by themselves.

\begin{figure}[h]
    \centering
    \includegraphics[width=0.9\linewidth]{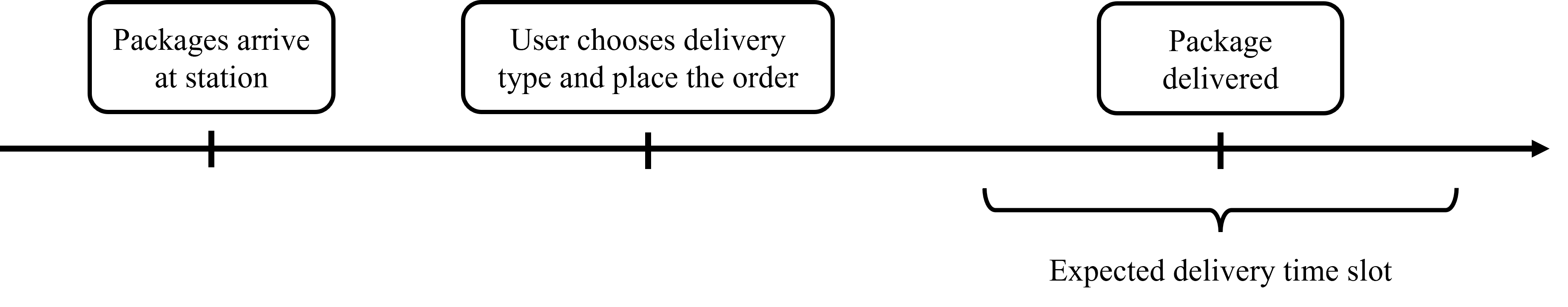}
    \caption{Timeline of delivery service}
    \label{fig:timeline}
\end{figure}

\begin{figure}[h]
    \centering
    \includegraphics[width=0.9\linewidth]{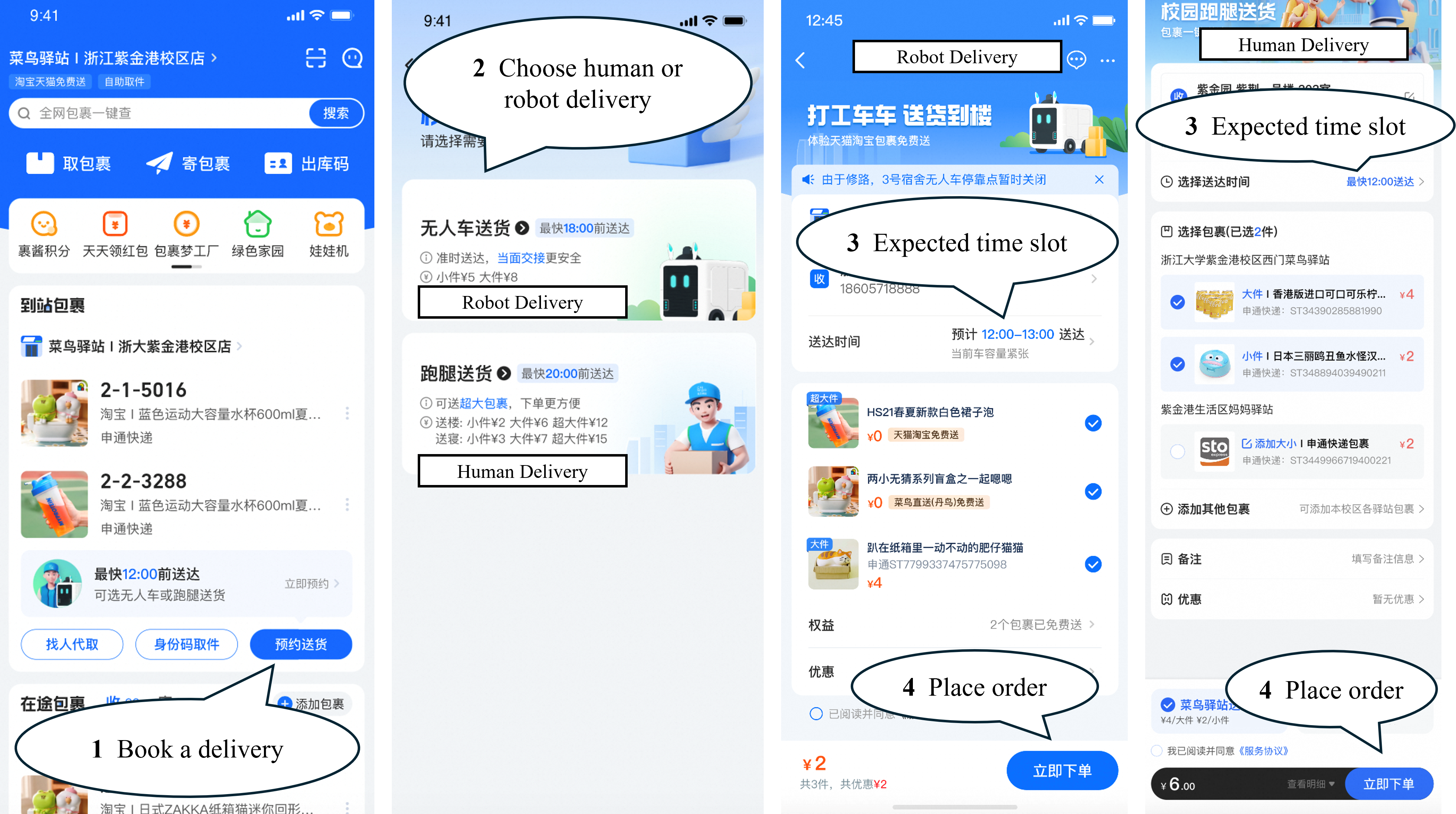}
    \caption{Ordering process of delivery services}
    \label{fig:ordering process}
\end{figure}

Our research context focuses on 32 Cainiao campus stations across China. These are the subset of campus stations that have included both delivery services during the year of 2024. After the inbounding of the packages, users can order a specific delivery service through which packages can be delivered from the station to a selected location. Figure \ref{fig:timeline} and Figure \ref{fig:ordering process} display the timeline of delivery services and the process of placing order respectively.


Figure \ref{fig:ordering process} illustrates the ordering process for last-mile delivery on Cainiao's mobile app, where users choose between paid human and robot delivery options. After a package arrives at the station, the user selects "Book a delivery", chooses a delivery type, picks a time window, and confirms the order. This process gives users control over both delivery method and timing, with price as a key consideration.

We collect package-level data from these 32 campus stations between January and November 2024. The dataset includes (i) delivery details (delivery type, fee, lead time), (ii) package attributes (product name, price, inbound time), and (iii) user demographics (gender, birthplace, device type), along with external variables such as weather, school level, and location. Out of 7,453,438 total package records, 241,517 involved either robot or human delivery, with the rest self-picked by users. Among delivered packages, 57\% were handled by robots and 43\% by human couriers, covering 61,630 unique users.

To test our hypotheses, we operationalize three key variables. First, sensitivity refers specifically and narrowly to adult/sexual products and contraceptive supplies (\textit{e.g.}, sex toys, condoms, and lingerie), where privacy concerns are most salient. Second, package value captures the listed price of the goods. Third, we use bad weather as a proxy for environmental complexity. Specifically, we define a weather condition as "bad" if it meets either of the following criteria: (1) a temperature warning is issued by the administration ($\geqslant$ 37 $^{\circ}$C or $\leqslant$ -6 $^{\circ}$C), or (2) any form of precipitation occurs, including rain and snow.



\section{Main Results} \label{sec:main results}
\subsection{Station-level Evidence}
We first construct a station-level panel to track the overall effect of sensitivity. We use the following OLS model with two-way fixed effect to estimate the station-level effect:
\begin{align}\label{eq1}
    type\_ratio_{st} = \alpha + \beta_1 \cdot sensitive\_ratio_{st} +\beta_0\cdot X_{st} + StationFE_s + MonthFE_t + \varepsilon_{st} 
\end{align}
where $type\_ratio_{st}$ denotes the percentage of robot (human) delivery over total package number of station $s$ in month $t$. $X_{st}$ are control variables. $StationFE_s$ and $MonthFE_t$ are station fixed effect and time fixed effect respectively. 

\begin{table}[ht!]

    \centering
    \footnotesize
    \caption{Station-level evidence: effect of sensitivity}

    \begin{tabular}{rcc}
\hline
                          & \begin{tabular}[c]{@{}c@{}}(1) \\ $ratio\_robot$\end{tabular}   & \begin{tabular}[c]{@{}c@{}}(2) \\ $ratio\_human$\end{tabular} \\ \hline
$const$                   & \begin{tabular}[c]{@{}c@{}}0.0106\\ (0.0077)\end{tabular}    & \begin{tabular}[c]{@{}c@{}}0.0104\\ (0.0063)\end{tabular}  \\
$sensitive\_ratio$        & \begin{tabular}[c]{@{}c@{}}1.1663**\\ (0.5518)\end{tabular}  & \begin{tabular}[c]{@{}c@{}}0.3020\\ (0.4515)\end{tabular}  \\
$\lg(avg\_package\_value)$ & \begin{tabular}[c]{@{}c@{}}3.711e-06\\ (0.0012)\end{tabular} & \begin{tabular}[c]{@{}c@{}}0.0010\\ (0.0010)\end{tabular}  \\
\multirow{2}{*}{$StationFE$}  & \multirow{2}{*}{Yes}    & \multirow{2}{*}{Yes}    \\
                              &                         &                         \\
\multirow{2}{*}{$TimeFE$}     & \multirow{2}{*}{Yes}    & \multirow{2}{*}{Yes}    \\
                              &                         &                         \\
\multirow{2}{*}{$R^2$}        & \multirow{2}{*}{0.0299} & \multirow{2}{*}{0.0094} \\
                              &                         &                         \\
\multirow{2}{*}{Observations} & \multirow{2}{*}{172}    & \multirow{2}{*}{172}    \\
                              &                         &                         \\ \hline
\multicolumn{3}{r}{{\small\textit{Note.} ** $p<$0.05, *** $p<$0.01. Standard errors are in parentheses.}}                                            
\end{tabular}

\label{tab:station-level evidence}

\end{table} 

Table \ref{tab:station-level evidence} presents the estimated result based on equation (\ref{eq1}). The ratio of sensitive packages has an significantly positive effect on the ratio robot delivery. When the ratio of sensitive packages increases by 1\%, number of packages delivered by robot vehicle will then increase by 1.17\%. The impact of package value does not show a significant effect on average at the station level. Weather effect is absorbed by station and time fixed effects. However, such significant effect is not observed in human delivery. To further investigate the impact on user choice behavior, we conduct the package-level analysis in the following section.

\subsection{Main Effects}

To empirically test the three hypotheses outlined above, we estimate an OLS model using the full delivery dataset of 241,517 package-level observations. The dependent variable $delivery\_type$ is a binary indicator equal to 1 if the delivery was completed by a robot and 0 if completed by a human courier. In addition, we incorporate both user fixed effects and time fixed effects to control for temporal variation in delivery conditions. Formally, our empirical model specification is as follows:

\begin{align*}\label{eq2}
    delivery\_type_{ijst} = \alpha + \beta_1 \cdot is\_package\_sensitive_{i} + \beta_2\cdot \lg(package\_value)_{i} \\+ \beta_3\cdot is\_bad\_weather_{st} +\beta_0\cdot X_{ijst} + TimeFE_t + \varepsilon_{ijst} \tag{2}
\end{align*}
where $delivery\_type_{ijst}$ is the binary choice of user $j$ for package $i$ in station $s$ at time $t$, where $delivery\_type_{ijst}$ equals $1$ for robot delivery and $0$ for human delivery. $is\_package\_sensitive_{i}$ is also a binary variable such that 1 for sensitive packages and 0 otherwise. $\lg(package\_value)_{i}$ denotes the log of the value of package $i$ that user $j$ paid. $is\_bad\_weather_{st}$ is another binary variable that represents the weather of station $s$ is bad or not at time $t$. $X_{ijst}$ are control variables consisting of three types of covariates: delivery information, user demographic and station-level variables. Following the setup in \textcite{amorim2024customer}, we include $lead\_time$, $lead\_time\_other$ and $delivery\_price$ to control the delivery-related effectiveness. To take into account demographic effect, gender ($is\_female$), device model ($is\_ios$), delivery time sensitivity ($is\_time\_sensitive$), and place of birth ($is\_South$) are included as well. Additionally, the station-level characteristics are also considered by incorporating school type ($is\_211$, "211 universities", China's equivalent of R1 universities in the US) and school location  ($is\_school\_South$).  $TimeFE_t$ considers the time fixed effect. 

\begin{table}[ht!]
\centering
\footnotesize
\caption{Main effects on user-level choice of delivery}
\label{tab:main result}
\begin{tabular}{rcccc}
\hline
\multicolumn{1}{l}{\multirow{2}{*}{}} &
  \multicolumn{4}{c}{\multirow{2}{*}{Dependent Variable: $delivery\_type$}} \\
\multicolumn{1}{l}{} &
  \multicolumn{4}{c}{} \\ \cline{2-5} 
\multirow{2}{*}{} &
  \multirow{2}{*}{(1)} &
  \multirow{2}{*}{(2)} &
  \multirow{2}{*}{(3)} &
  \multirow{2}{*}{(4)} \\
 &
   &
   &
   &
   \\ \hline
$const$ &
  \begin{tabular}[c]{@{}c@{}}0.5473*** \\ (0.0034)\end{tabular} &
  \begin{tabular}[c]{@{}c@{}}0.8596*** \\ (0.0037)\end{tabular} &
  \begin{tabular}[c]{@{}c@{}}0.8920*** \\ (0.0039)\end{tabular} &
  \begin{tabular}[c]{@{}c@{}}0.9069*** \\ (0.0040)\end{tabular} \\
$is\_package\_sensitive$ &
  \begin{tabular}[c]{@{}c@{}}0.1392*** \\ (0.0158)\end{tabular} &
  \begin{tabular}[c]{@{}c@{}}0.1224*** \\ (0.0146)\end{tabular} &
  \begin{tabular}[c]{@{}c@{}}0.1095*** \\ (0.0146)\end{tabular} &
  \begin{tabular}[c]{@{}c@{}}0.1149*** \\ (0.0143)\end{tabular} \\
$\lg(package\_value)$ &
  \begin{tabular}[c]{@{}c@{}}0.0059*** \\ (0.0008)\end{tabular} &
  \begin{tabular}[c]{@{}c@{}}0.0127*** \\ (0.0007)\end{tabular} &
  \begin{tabular}[c]{@{}c@{}}0.0135*** \\ (0.0007)\end{tabular} &
  \begin{tabular}[c]{@{}c@{}}0.0097*** \\ (0.0007)\end{tabular} \\
$is\_bad\_weather$ &
  \begin{tabular}[c]{@{}c@{}}-0.0135*** \\ (0.0024)\end{tabular} &
  \begin{tabular}[c]{@{}c@{}}-0.0426*** \\ (0.0022)\end{tabular} &
  \begin{tabular}[c]{@{}c@{}}-0.0398*** \\ (0.0022)\end{tabular} &
  \begin{tabular}[c]{@{}c@{}}-0.0163*** \\ (0.0022)\end{tabular} \\
$lead\_time$ &
   &
  \begin{tabular}[c]{@{}c@{}}-0.0241*** \\ (0.0002)\end{tabular} &
  \begin{tabular}[c]{@{}c@{}}-0.0239*** \\ (0.0002)\end{tabular} &
  \begin{tabular}[c]{@{}c@{}}-0.0235*** \\ (0.0002)\end{tabular} \\
$lead\_time\_other$ &
   &
  \begin{tabular}[c]{@{}c@{}}0.0040*** \\ (0.0002)\end{tabular} &
  \begin{tabular}[c]{@{}c@{}}0.0040*** \\ (0.0002)\end{tabular} &
  \begin{tabular}[c]{@{}c@{}}0.0038*** \\ (0.0002)\end{tabular} \\
$delivery\_price$ &
   &
  \begin{tabular}[c]{@{}c@{}}-0.1628*** \\ (0.0010)\end{tabular} &
  \begin{tabular}[c]{@{}c@{}}-0.1658*** \\ (0.0010)\end{tabular} &
  \begin{tabular}[c]{@{}c@{}}-0.1489*** \\ (0.0010)\end{tabular} \\
$is\_female$ &
   &
   &
  \begin{tabular}[c]{@{}c@{}}-0.0449***\\ (0.0020)\end{tabular} &
  \begin{tabular}[c]{@{}c@{}}-0.0590*** \\ (0.0019)\end{tabular} \\
$is\_ios$ &
   &
   &
  \begin{tabular}[c]{@{}c@{}}0.0064*** \\ (0.0019)\end{tabular} &
  \begin{tabular}[c]{@{}c@{}}0.0047** \\ (0.0018)\end{tabular} \\
$is\_time\_sensitive$ &
   &
   &
  \begin{tabular}[c]{@{}c@{}}0.0040 \\ (0.0025)\end{tabular} &
  \begin{tabular}[c]{@{}c@{}}0.0020\\ (0.0024)\end{tabular} \\
$is\_South$ &
   &
   &
  \begin{tabular}[c]{@{}c@{}}-0.0306*** \\ (0.0020)\end{tabular} &
  \begin{tabular}[c]{@{}c@{}}0.0178*** \\ (0.0021)\end{tabular} \\
$is\_school\_South$ &
   &
   &
   &
  \begin{tabular}[c]{@{}c@{}}-0.1140*** \\ (0.0022)\end{tabular} \\
$is\_211$ &
   &
   &
   &
  \begin{tabular}[c]{@{}c@{}}0.2101*** \\ (0.0026)\end{tabular} \\
\multirow{2}{*}{$TimeFE$} &
  \multirow{2}{*}{Yes} &
  \multirow{2}{*}{Yes} &
  \multirow{2}{*}{Yes} &
  \multirow{2}{*}{Yes} \\
 &
   &
   &
   &
   \\
\multirow{2}{*}{$R^2$} &
  \multirow{2}{*}{0.0007} &
  \multirow{2}{*}{0.1401} &
  \multirow{2}{*}{0.1435} &
  \multirow{2}{*}{0.1732} \\
 &
   &
   &
   &
   \\
\multirow{2}{*}{Observations} &
  \multirow{2}{*}{241,517} &
  \multirow{2}{*}{241,517} &
  \multirow{2}{*}{241,517} &
  \multirow{2}{*}{241,517} \\
 &
   &
   &
   &
   \\ \hline
\multicolumn{5}{l}{\small\textit{Note.} *$p<$0.1, ** $p<$0.05, *** $p<$0.01. Standard errors are in parentheses.}
\end{tabular}
\end{table} 

Table \ref{tab:main result}  demonstrates our main result. Following the approach of \textcite{chen2013effect}, we first estimate the main effects of interest. Column (1) shows direct estimation towards our hypotheses. Column (2) to (4) incorporate delivery covariates, demographic characteristics and station-level characteristics respectively step-by-step. Column (4) presents our final estimation.

We empirically test the three hypotheses proposed earlier. Consistent with H1, we find that consumers are significantly more likely to choose robot delivery when the package is privacy-sensitive. As shown in Table \ref{tab:main result}, the probability of selecting robot delivery increases by 11.49\% ($p<0.001$) for sensitive items. This sizable effect validates our proposition that privacy concerns, particularly the desire to avoid social judgment, drive users toward a less socially intrusive service type, making robot delivery the preferred option in such contexts.

Regarding H2, our results support the mechanism outlined in H2b. We find that a 1\% increase in package value is associated with a 0.97\% increase in the likelihood of choosing robot delivery ($p<0.001$). This effect suggests that, consumers view robots as sufficiently reliable and secure to handle valuable goods, likely due to their perceived consistency and traceability. 

Finally, in support of H3, we find that bad weather conditions significantly increase the preference for human delivery. Specifically, under conditions of extreme temperature or precipitation, the probability of choosing robot delivery decreases by 1.63\% ($p<0.001$), implying a corresponding increase in preference for human couriers. This finding aligns with our hypothesis that environmental complexity heightens trust expectations, prompting users to rely on humans for their adaptive judgment and contextual flexibility.

Taken together, these empirical results directly address our central research question regarding how consumers make service type choices in human-robot coexistence settings. By demonstrating that delivery service choices are systematically influenced by contextual triggers through psychologically grounded mechanisms, our findings validate the theoretical framework developed earlier. This evidence highlights the nuanced and situational nature of consumer decision-making in technology-mediated services. In doing so, we contribute to a deeper understanding of the transitional phase of human-robot service coexistence and provide a behavioral foundation for designing more adaptive, context-aware service systems.

\section{Heterogeneous Analysis} \label{heterogeneous analysis}

Building upon our baseline findings, we further investigate how user-level traits moderate the effects of three drivers on delivery service choices. Specifically, we explore two forms of heterogeneity: (1) gender differences and (2) the role of user experience, focusing on behavioral adaptation through repeated interactions with robot and human delivery services. In addition, we also model user's habit formation and reactions to price adjustment in this section.

\subsection{Impact of Gender Heterogeneity}
To examine gender differences in delivery preferences, we interact the three contextual drivers with a gender indicator variable ($is\_female$). The gender heterogeneous model is as follows.

\begin{align*}
    \label{eq3}
    delivery\_type_{ijst} =&\ \alpha + \beta_1 \cdot is\_sensitive_{i} + \beta_2\cdot \lg(package\_value)_{i} + \beta_3\cdot is\_bad\_weather_{st}\\ 
    & +\gamma_1\cdot(is\_sensitive_{i}\times is\_female_j)\\
    &+\gamma_2\cdot(\lg(package\_value)_{i}\times is\_female_j)\\
    &+\gamma_3\cdot(is\_bad\_weather_{st}\times is\_female_j)\\
    &+\beta_0\cdot X_{ijst} + TimeFE_t + \varepsilon_{ijst} \tag{3}
\end{align*}

\begin{table}[ht!]
\centering
\caption{Heterogeneity effect of gender and user historical behavior}
\scriptsize
\label{tab:hte result}
\begin{tabular}{rccc}
\hline
\multicolumn{1}{l}{\multirow{2}{*}{}} &
  \multicolumn{3}{c}{\multirow{2}{*}{Dependent Variable: $delivery\_type$}} \\
\multicolumn{1}{l}{} &
  \multicolumn{3}{c}{} \\ \cline{2-4} 
\multirow{2}{*}{} &
  \multirow{2}{*}{(1)} &
  \multirow{2}{*}{(2)} &
  \multirow{2}{*}{(3)} \\
 &
   &
   &
   \\ \hline
$const$ &
  \begin{tabular}[c]{@{}c@{}}0.8963*** \\ (0.0053)\end{tabular} &
  \begin{tabular}[c]{@{}c@{}}0.8347***\\ (0.0032)\end{tabular} &
  \begin{tabular}[c]{@{}c@{}}0.8396***\\ (0.0031)\end{tabular} \\
$is\_package\_sensitive$ &
  \begin{tabular}[c]{@{}c@{}}0.1285*** \\ (0.0173)\end{tabular} &
  \begin{tabular}[c]{@{}c@{}}0.0635***\\ (0.0111)\end{tabular} &
  \begin{tabular}[c]{@{}c@{}}0.0644***\\ (0.0104)\end{tabular} \\
$is\_package\_sensitive \times is\_female$ &
  \begin{tabular}[c]{@{}c@{}}-0.0463 \\ (0.0310)\end{tabular} &
   &
   \\
\multicolumn{1}{l}{$is\_package\_sensitive \times robot\_cumu\_num$} &
   &
  \begin{tabular}[c]{@{}c@{}}-0.0005\\ (0.0011)\end{tabular} &
   \\
$is\_package\_sensitive \times human\_cumu\_num$ &
   &
   &
  \begin{tabular}[c]{@{}c@{}}-0.0010\\ (0.0007)\end{tabular} \\
$\lg(package\_value)$ &
  \begin{tabular}[c]{@{}c@{}}0.0116*** \\ (0.0011)\end{tabular} &
  \begin{tabular}[c]{@{}c@{}}0.0061***\\ (0.0005)\end{tabular} &
  \begin{tabular}[c]{@{}c@{}}0.0050***\\ (0.0005)\end{tabular} \\
$\lg(package\_value)\times is\_female$ &
  \begin{tabular}[c]{@{}c@{}}-0.0032** \\ (0.0014)\end{tabular} &
   &
   \\
$\lg(package\_value) \times robot\_cumu\_num$ &
   &
  \begin{tabular}[c]{@{}c@{}}-0.0003***\\ (0.0000)\end{tabular} &
   \\
$\lg(package\_value) \times human\_cumu\_num$ &
   &
   &
  \begin{tabular}[c]{@{}c@{}}-0.0000\\ (0.0000)\end{tabular} \\
$is\_bad\_weather$ &
  \begin{tabular}[c]{@{}c@{}}-0.0084*** \\ (0.0031)\end{tabular} &
  \begin{tabular}[c]{@{}c@{}}-0.0159***\\ (0.0017)\end{tabular} &
  \begin{tabular}[c]{@{}c@{}}-0.0166***\\ (0.0016)\end{tabular} \\
$is\_bad\_weather\times is\_female$ &
  \begin{tabular}[c]{@{}c@{}}-0.0138*** \\ (0.0037)\end{tabular} &
   &
   \\
$is\_bad\_weather \times robot\_cumu\_num$ &
   &
  \begin{tabular}[c]{@{}c@{}}0.0001\\ (0.0001)\end{tabular} &
   \\
$is\_bad\_weather \times human\_cumu\_num$ &
   &
   &
  \begin{tabular}[c]{@{}c@{}}0.0004***\\ (0.0001)\end{tabular} \\
$lead\_time$ &
  \begin{tabular}[c]{@{}c@{}}-0.0235*** \\ (0.0002)\end{tabular} &
  \begin{tabular}[c]{@{}c@{}}-0.0102***\\ (0.0002)\end{tabular} &
  \begin{tabular}[c]{@{}c@{}}-0.0102***\\ (0.0002)\end{tabular} \\
$lead\_time\_other$ &
  \begin{tabular}[c]{@{}c@{}}0.0038*** \\ (0.0002)\end{tabular} &
  \begin{tabular}[c]{@{}c@{}}0.0009***\\ (0.0001)\end{tabular} &
  \begin{tabular}[c]{@{}c@{}}0.0009***\\ (0.0001)\end{tabular} \\
$delivery\_price$ &
  \begin{tabular}[c]{@{}c@{}}-0.1489*** \\ (0.0010)\end{tabular} &
  \begin{tabular}[c]{@{}c@{}}-0.0945***\\ (0.0007)\end{tabular} &
  \begin{tabular}[c]{@{}c@{}}-0.0945***\\ (0.0007)\end{tabular} \\
$is\_female$ &
  \begin{tabular}[c]{@{}c@{}}-0.0403*** \\ (0.0063)\end{tabular} &
  \begin{tabular}[c]{@{}c@{}}-0.0225***\\ (0.0014)\end{tabular} &
  \begin{tabular}[c]{@{}c@{}}-0.0226***\\ (0.0014)\end{tabular} \\
$is\_ios$ &
  \begin{tabular}[c]{@{}c@{}}0.0047** \\ (0.0018)\end{tabular} &
  \begin{tabular}[c]{@{}c@{}}0.0024*\\ (0.0013)\end{tabular} &
  \begin{tabular}[c]{@{}c@{}}0.0024*\\ (0.0013)\end{tabular} \\
$is\_time\_sensitive$ &
  \begin{tabular}[c]{@{}c@{}}0.0020 \\ (0.0024)\end{tabular} &
  \begin{tabular}[c]{@{}c@{}}-0.0003\\ (0.0017)\end{tabular} &
  \begin{tabular}[c]{@{}c@{}}-0.0001\\ (0.0017)\end{tabular} \\
$is\_South$ &
  \begin{tabular}[c]{@{}c@{}}0.0178*** \\ (0.0021)\end{tabular} &
  \begin{tabular}[c]{@{}c@{}}0.0157***\\ (0.0015)\end{tabular} &
  \begin{tabular}[c]{@{}c@{}}0.0157***\\ (0.0015)\end{tabular} \\
$is\_school\_South$ &
  \begin{tabular}[c]{@{}c@{}}-0.1142*** \\ (0.0022)\end{tabular} &
  \begin{tabular}[c]{@{}c@{}}-0.0570***\\ (0.0016)\end{tabular} &
  \begin{tabular}[c]{@{}c@{}}-0.0569***\\ (0.0016)\end{tabular} \\
$is\_211$ &
  \begin{tabular}[c]{@{}c@{}}0.2103*** \\ (0.0026)\end{tabular} &
  \begin{tabular}[c]{@{}c@{}}0.0449***\\ (0.0020)\end{tabular} &
  \begin{tabular}[c]{@{}c@{}}0.0449***\\ (0.0020)\end{tabular} \\
$robot\_cumu\_num$ &
   &
  \begin{tabular}[c]{@{}c@{}}0.0019***\\ (0.0002)\end{tabular} &
  \begin{tabular}[c]{@{}c@{}}0.0007*** \\ (0.0001)\end{tabular} \\
$human\_cumu\_num$ &
   &
  \begin{tabular}[c]{@{}c@{}}-0.0016*** \\ (0.0000)\end{tabular} &
  \begin{tabular}[c]{@{}c@{}}-0.0018*** \\ (0.0001)\end{tabular} \\
$last\_choice\_robot$ &
   &
  \begin{tabular}[c]{@{}c@{}}0.2456*** \\ (0.0017)\end{tabular} &
  \begin{tabular}[c]{@{}c@{}}0.2461*** \\ (0.0017)\end{tabular} \\
$last\_choice\_human$ &
   &
  \begin{tabular}[c]{@{}c@{}}-0.4967*** \\ (0.0018)\end{tabular} &
  \begin{tabular}[c]{@{}c@{}}-0.4966*** \\ (0.0018)\end{tabular} \\
\multirow{2}{*}{$TimeFE$} &
  \multirow{2}{*}{Yes} &
  \multirow{2}{*}{Yes} &
  \multirow{2}{*}{Yes} \\
 &
   &
   &
   \\
\multirow{2}{*}{$R^2$} &
  \multirow{2}{*}{0.1733} &
  \multirow{2}{*}{0.5756} &
  \multirow{2}{*}{0.5755} \\
 &
   &
   &
   \\
\multirow{2}{*}{Observations} &
  \multirow{2}{*}{241,517} &
  \multirow{2}{*}{241,517} &
  \multirow{2}{*}{241,517} \\
 &
   &
   &
   \\ \hline
\multicolumn{4}{l}{\small\textit{Note.} *$p<$0.1, ** $p<$0.05, *** $p<$0.01. Standard errors are in parentheses.}
\end{tabular}
\end{table}

As shown in Column (1) of Table \ref{tab:hte result}, while all three main effects remain significant and directionally consistent with our earlier findings, we observe notable gender-based moderation in two contexts. Specifically, the interaction between bad weather and female user is significantly negative (-0.0138, $p$ < 0.01), indicating that female users are even more likely than male users to prefer human delivery under adverse environmental conditions. This result suggests that women may place greater trust in human couriers when facing uncertainty, aligning with prior literature on gender differences in risk perception and trust preferences. Conversely, the interactions between gender and the effects of privacy sensitivity and product value (-0.0032, $p$ < 0.05) are weaker. The latter suggests that female users are slightly less likely to choose robot delivery as product value increases, potentially reflecting more cautious attitudes toward automation in high-stakes scenarios. We do not find the interaction effect between gender and package sensitivity. One possible explanation is that we adopt a very narrow definition of sensitive packages, and both male and female users may experience social embarrassment when receiving sexual products or contraceptive supplies.

\subsection{Interaction Effect of Main Factors}

\begin{table}[ht!]
\centering
\caption{Interaction effect of main factors}
\label{tab:interaction effect}
\footnotesize
\begin{tabular}{rc}
\hline
\multirow{2}{*}{}                                  & \multicolumn{1}{l}{\multirow{2}{*}{Dependent Variable: $delivery\_type$}} \\
                              & \multicolumn{1}{l}{}                                          \\ \cline{2-2} 
\multirow{2}{*}{\textbf{}}    & \multirow{2}{*}{(1)}                                          \\
                              &                                                               \\ \hline
$const$                       & \begin{tabular}[c]{@{}c@{}}-0.0962***\\ (0.0046)\end{tabular} \\
$is\_package\_sensitive \times lg\_package\_value$ & \begin{tabular}[c]{@{}c@{}}-0.0211\\ (0.0141)\end{tabular}                \\
$is\_package\_sensitive \times is\_bad\_weather$   & \begin{tabular}[c]{@{}c@{}}0.0637**\\ (0.0292)\end{tabular}               \\
$lg\_package\_value \times is\_bad\_weather$       & \begin{tabular}[c]{@{}c@{}}-0.0019\\ (0.0014)\end{tabular}                \\
$is\_package\_sensitive$                           & \begin{tabular}[c]{@{}c@{}}0.1874***\\ (0.0691)\end{tabular}              \\
$lg\_package\_value$          & \begin{tabular}[c]{@{}c@{}}0.0105***\\ (0.0009)\end{tabular}  \\
$is\_bad\_weather$            & \begin{tabular}[c]{@{}c@{}}-0.0090\\ (0.0062)\end{tabular}    \\
$lead\_time$                  & \begin{tabular}[c]{@{}c@{}}-0.0235***\\ (0.0002)\end{tabular} \\
$lead\_time\_other$           & \begin{tabular}[c]{@{}c@{}}0.0038***\\ (0.0002)\end{tabular}  \\
$delivery\_price$             & \begin{tabular}[c]{@{}c@{}}-0.1489***\\ (0.0010)\end{tabular} \\
$is\_female$                  & \begin{tabular}[c]{@{}c@{}}-0.0590***\\ (0.0019)\end{tabular} \\
$is\_ios$                     & \begin{tabular}[c]{@{}c@{}}0.0047**\\ (0.0018)\end{tabular}   \\
$is\_time\_sensitive$         & \begin{tabular}[c]{@{}c@{}}0.0020\\ (0.0024)\end{tabular}     \\
$is\_South$                   & \begin{tabular}[c]{@{}c@{}}0.0178***\\ (0.0021)\end{tabular}  \\
$is\_school\_South$           & \begin{tabular}[c]{@{}c@{}}-0.1140***\\ (0.0022)\end{tabular} \\
$is\_211$                     & \begin{tabular}[c]{@{}c@{}}0.2101***\\ (0.0026)\end{tabular}  \\
\multirow{2}{*}{$TimeFE$}     & \multirow{2}{*}{Yes}                                          \\
                              &                                                               \\
\multirow{2}{*}{$R^2$}        & \multirow{2}{*}{0.1732}                                       \\
                              &                                                               \\
\multirow{2}{*}{Observations} & \multirow{2}{*}{241,517}                                      \\
                              &                                                               \\ \hline
\multicolumn{2}{l}{\small\textit{Note.} * $p<$0.1, ** $p<$0.05, *** $p<$0.01. Standard errors are in parentheses.}            
\end{tabular}
\end{table}

We next examine whether contextual factors jointly influence delivery choice. Rather than treating three main factors independently, we explore their interaction: for example, a privacy-sensitive item delivered during rainy day may trigger qualitatively different preferences than would be predicted by each condition alone. Table \ref{tab:interaction effect} shows a significant interaction between $is\_package\_sensitive$ and $is\_bad\_weather$ (0.0637, $p<0.05$), indicating that users tend to choose robot delivery for sensitive packages when the weather is bad.

The main effect estimates show two countervailing forces: $is\_package\_sensitive$ increases the probability of selecting robot, whereas $is\_bad\_weather$ decreases it.   However, bad weather conditions can attenuate the effect of sensitive packages (\textit{i.e.}, coefficient of the interaction term is smaller than that of $is\_package\_sensitive$ alone in main effect estimates). More specifically, while sensitive packages tend to push customers toward robots, that push is weakened in bad weather. Customers trade off privacy concerns against weather-related concerns. This pattern is consistent with theories that distinguish social-evaluative drivers (which favor robots for sensitive consumption) from competence drivers (which favor humans under uncertainty), which is reflected in \textcite{holthower2023robots} on embarrassment reduction and \textcite{Wirtz2018} on trust and adaptability.

\subsection{Modeling Dynamic Preferences and Habit Formation}

To examine whether consumer delivery preferences evolve over time, we extend our baseline specification to incorporate dynamic behavioral factors such as habit formation, trust accumulation, and contextual recalibration. Specifically, to assess whether repeated experience with robot or human delivery influences subsequent choices, we augment our model with dynamic behavioral indicators. These include whether a user's last choice was handled by a robot or human (the user is a new customer when both are zero), as well as the cumulative number of each delivery type. We augment our model as follows.
{ \small
\begin{align*} \label{eq4} 
delivery\_type_{ijst} =&\ \alpha + \beta_1 \cdot is\_package\_sensitive_{i} + \beta_2\cdot \lg(package\_value)_{i} + \beta_3\cdot is\_bad\_weather_{st}\\ 
&+ \delta_1 \cdot last\_choice\_robot_{jt} + \delta_2 \cdot last\_choice\_human_{jt}\\
&+ \beta_4 \cdot robot\_cumu\_num_{jt} + \beta_5 \cdot human\_cumu\_num_{jt} \\
&+ \gamma_1 \cdot (is\_sensitive_{i} \times robot(human)\_cumu\_num_{jt}) \\
&+ \gamma_2 \cdot (\log(package\_value)_{i} \times robot(human)\_cumu\_num_{jt}) \\
&+ \gamma_3 \cdot (is\_bad\_weather_{st} \times robot(human)\_cumu\_num_{jt}) \\
&+ \beta_0 \cdot X_{ijst} + TimeFE_t + \varepsilon_{ijst} \tag{4}
\end{align*}
}%

\noindent where $last\_choice\_robot(human)_{jt}$ indicates whether user $j$ chose robot(human) delivery on their most recent prior transaction. $robot(human)\_cumu\_num_{jt}$ captures the cumulative number of robot(human) deliveries made by user $j$ up to time $t$ (current order not included). The interaction coefficients $\gamma_1$ -- $\gamma_3$ test whether the main effects are moderated by accumulated robot(human) service experience.

As reported in Columns (2) and (3) of Table \ref{tab:hte result}, behavioral history significantly shapes current choices. Users are more likely to repeat their previous choice: having selected robot delivery in the prior transaction increases the likelihood of choosing it again by 24\% -- 25\% ($p$ < 0.001), while choosing a human service in the previous choice reduces the probability of selecting a robot in the current round by around 49.7\% ($p$ < 0.001). These strong path dependencies suggest substantial habit formation in delivery service decisions. Further, cumulative experience also plays a role: the coefficient on $robot\_cumu\_num$ is positive and significant (0.0007--0.0019, $p$ < 0.001), while that on $human\_cumu\_num$ is negative and significant (-0.0018 -- -0.0016, $p$ < 0.001), indicating that growing familiarity with a delivery type reinforces user preference for that type.

Interestingly, as users accumulate more experience with robot delivery, the positive effect of product value on the likelihood of choosing a robot diminishes ($\lg(package\_value) \times robot\_cumu\_num$: -0.0003, $p<0.001$). While first-time or infrequent robot users are more inclined to choose robots for high-value packages -- likely due to perceptions of security and traceability -- this sensitivity weakens over time. Experienced users may grow more confident in robot performance across all product categories, and as a result, no longer anchor their delivery choice to item value. 

On the other hand, more frequent use of human service also softens the drop in robot delivery preference under adverse weather conditions ($is\_bad\_weather \times human\_cumu\_num$: 0.0004, $p<0.001$). In the baseline, bad weather discourages robot use due to concerns about mechanical failure or limited adaptability. However, users with higher human delivery experience are less reactive to such trust concern: they are less likely to switch away from robots when facing environmental uncertainty. One possible interpretation is the relative trust recalibration: frequent human delivery users may realize the practical limits of human adaptability or experience inconsistent performance in poor weather, which makes robot delivery comparatively more acceptable over time. 

This dynamic specification enables us to move beyond static heterogeneity and uncover the trajectory of consumer adaptation in hybrid human-robot service systems.

\subsection{Reactions to Dynamic Delivery Pricing}
We further investigate the impact of delivery price on user preference, which makes it possible to estimate the elasticity of the delivery price. During the study period, four stations adjust their prices for either of the delivery types.  Two stations (station 1 and station 2) decreased price of human service by 1 RMB (from 2 RMB to 1 RMB) and 1 RMB (from 1 RMB to free) respectively. Two stations (station 3 and station 4) increased price of robot service by 0.6 RMB (from 2.2 RMB to 2.8 RMB) and 1 RMB (from 1.5 RMB to 2.5 RMB) respectively. Price changing is exogenous to users. We model this price adjustment by a Difference-in-Difference (DiD) model as follows.
\begin{align*}  \label{eq5}
    delivery\_type_{ijst} = 
    & \alpha\\
    & + \beta \cdot after\_human(robot)\_price\_change_{st} \times is\_human(robot)\_price\_change_s \\
    & + \gamma \cdot X_{ijst}\\
    & + UserFE_j + TimeFE_t + \varepsilon_{ijst} \tag{5}
\end{align*}

\noindent where $after\_human(robot)\_price\_change_{st}$ equals 1 if, for station $s$, period $t$ falls after that station's price change for the corresponding delivery type (human or robot), and 0 otherwise; $is\_human(robot)\_price\_change_s$ equals 1 for stations that ever changed the corresponding type's price during the sample window (treated stations), and 0 for stations that never did (controls). We estimate the specification separately for human-fee changes and robot-fee changes; in each case, the interaction $after \times is$ is the DiD term, and its coefficient $\beta$ identifies the average treatment effect on the treated (ATT).

\begin{table}[ht!]
\centering
\caption{DiD Result for Price Changing}
\footnotesize
\label{tab:did for price changing}
\begin{tabular}{rcccc}
\hline
\multicolumn{1}{l}{\multirow{2}{*}{}} &
  \multicolumn{4}{c}{\multirow{2}{*}{Dependent Variable: $delivery\_type$}} \\
\multicolumn{1}{l}{} &
  \multicolumn{4}{c}{} \\ \cline{2-5} 
\textbf{} &
  \begin{tabular}[c]{@{}c@{}}(1)\\ Station 1\end{tabular} &
  \begin{tabular}[c]{@{}c@{}}(2)\\ Station 2\end{tabular} &
  \begin{tabular}[c]{@{}c@{}}(3)\\ Station 3\end{tabular} &
  \begin{tabular}[c]{@{}c@{}}(4)\\ Station 4\end{tabular} \\ \hline
$const$ &
  \begin{tabular}[c]{@{}c@{}}0.5863***\\ (0.2237)\end{tabular} &
  \begin{tabular}[c]{@{}c@{}}0.5697**\\ (0.2240)\end{tabular} &
  \begin{tabular}[c]{@{}c@{}}0.5694**\\ (0.2241)\end{tabular} &
  \begin{tabular}[c]{@{}c@{}}0.5675**\\ (0.2241)\end{tabular} \\
\begin{tabular}[c]{@{}r@{}}$after\_human\_price\_decrease\_station1$\\ $\times\; is\_human\_price\_decrease\_station1$\end{tabular} &
  \begin{tabular}[c]{@{}c@{}}-0.2721***\\ (0.0116)\end{tabular} &
   &
   &
   \\
\begin{tabular}[c]{@{}r@{}}$after\_human\_price\_decrease\_station2$\\ $\times\; is\_human\_price\_decrease\_station2$\end{tabular} &
   &
  \begin{tabular}[c]{@{}c@{}}-0.1974***\\ (0.0169)\end{tabular} &
   &
   \\
\begin{tabular}[c]{@{}r@{}}$after\_robot\_price\_increase\_station3$\\ $\times\; is\_robot\_price\_increase\_station3$\end{tabular} &
   &
   &
  \begin{tabular}[c]{@{}c@{}}-0.0133**\\ (0.0055)\end{tabular} &
   \\
\begin{tabular}[c]{@{}r@{}}$after\_robot\_price\_increase\_station4$\\ $\times\; is\_robot\_price\_increase\_station4$\end{tabular} &
   &
   &
   &
  \begin{tabular}[c]{@{}c@{}}-0.1469***\\ (0.0374)\end{tabular} \\
\multirow{2}{*}{$control\_variables$} &
  \multicolumn{4}{c}{\multirow{2}{*}{$Omitted$}} \\
 &
  \multicolumn{4}{c}{} \\
\multirow{2}{*}{$User FE$} &
  \multirow{2}{*}{Yes} &
  \multirow{2}{*}{Yes} &
  \multirow{2}{*}{Yes} &
  \multirow{2}{*}{Yes} \\
 &
   &
   &
   &
   \\
\multirow{2}{*}{$Time FE$} &
  \multirow{2}{*}{Yes} &
  \multirow{2}{*}{Yes} &
  \multirow{2}{*}{Yes} &
  \multirow{2}{*}{Yes} \\
 &
   &
   &
   &
   \\
\multirow{2}{*}{$R^2$} &
  \multirow{2}{*}{0.1387} &
  \multirow{2}{*}{0.1367} &
  \multirow{2}{*}{0.1361} &
  \multirow{2}{*}{0.1361} \\
 &
   &
   &
   &
   \\
   \hline
\multicolumn{5}{l}{\small\textit{Note.} *$p<$0.1, ** $p<$0.05, *** $p<$0.01. Standard errors are in parentheses.}
\end{tabular}
\end{table}

Table \ref{tab:did for price changing} shows the ATT observations with keeping others in the control. Either decreasing the price of human service or increasing the price of robot delivery intuitively makes users use more human service: ATT of station 1 is -0.2721 ($p<0.001$), for which the price of human service decreases from 2 RMB to 1 RMB; ATT of station 2 is -0.1974 ($p<0.001$), for which the price of human service decreases from 1 RMB to 0; ATT of station 3 is -0.0133 ($p<0.001$), for which the price of robot service increases from 2.2 RMB to 2.8 RMB; ATT of station 4 is -0.1469 ($p<0.001$), for which the price of robot service increases from 1.5 RMB to 2.5 RMB. 

Based on above estimation, we construct a semi-elasticity (SE), $semi\_elasticity_P = \frac{ATT_P}{\Delta P} \times 100\%$, reflecting user price sensitivity. The semi-elasticity measures the percentage change in preference of robot per RMB. Although the signs of SEs are the same, the absolute value are 27.21, 19.74, 2.22 and 14.69, respectively. Users are more price-sensitive for human service than robot service averagely.

\section{Robustness Check}  \label{robustness check}

\subsection{Alternative Choice Models}
We implement two different discrete regressions to strengthen our main results. As shown in Appendix B Table \ref{tab:logit and probit}, using Logit and Probit regressions does not change our findings fundamentally, where users tend to have their sensitive or high-value packages to be delivered by robots and order human delivery service under adverse weather conditions.

\subsection{User-level Fixed Effect}
We excluded user-level fixed effect in our main result due to the consideration to check the impact of user traits. In this section, we add user-level fixed effect to further examine the robustness. Appendix B Table \ref{tab:two-way effect} shows that sensitivity and package value remain significant but not for bad weather. Stations chosen in this study are campus-based and generally far away from each other. As a result, no packages for the same user were distributed to different stations at the same day. Because the variable $is\_bad\_weather$ describes station-wide bad weather on a day, adding two-way fixed effects will remove the variation that previously identified the coefficient (\textit{i.e.}, two-way fixed effects soak up the variation since $is\_bad\_weather$ is measured at station-daily level).

This section incorporates two additional robustness checks with more to be added.

\section{Conclusion} \label{sec:conclusion}
This paper utilizes 241,517 package-level delivery choices from 32 Cainiao stations to examine how product privacy sensitivity, product value, and environmental complexity shape consumer selections between human and robot last-mile delivery. To the best of our knowledge, this is the first empirical investigation of consumer decision-making in human-robot delivery service environments in a real-world setting. By uncovering how contextual factors and psychological mechanisms jointly shape delivery choices, our study not only contributes to the understanding of human-robot interaction in operational settings but also paves the way for future research on hybrid human-robot service system design, user heterogeneity, and the evolving role of automation in consumer-facing services.

\subparagraph{Main empirical findings.}
Our findings reveal that consumer preferences in hybrid human-robot service environments are systematically shaped by contextual factors that activate distinct psychological mechanisms. Consumers are more likely to choose robot delivery for privacy-sensitive and high-value products, reflecting a desire to reduce social exposure and a perception of robots as secure, traceable, and consistent. In contrast, adverse weather shifts preference toward human couriers, highlighting trust in human adaptability under uncertainty. In addition, we find that gender moderates these effects: female users are especially likely to switch to human delivery under bad weather. The interaction estimates show that bad weather attenuates the positive effect of privacy sensitivity. We also document pronounced behavioral dynamics: prior robot choice raises the likelihood of repeating robot delivery, while prior human choice reduces robot selection, and cumulative experience with a delivery type reinforces continued use (small but significant cumulative effects). These results demonstrate that delivery choices are driven not only by functional considerations such as price or speed, but also by deeper psychological concerns of users. Finally, we show the different price sensitivities of both delivery types respectively through DiD. 

\subparagraph{Theoretical contribution.}
These findings advance theory on robot service choice in three ways. First, they provide empirical evidence to show that consumer adoption of automation is not determined solely by instrumental trade-offs (price, speed) but also by context-activated psychological processes: 1) privacy sensitivity prompts an avoidance of socially evaluative human interaction, elevating the appeal of non-judgmental robots; 2) product value can amplify concerns about loss or accountability and, in monitored campus settings, can alternatively increase trust in robots' traceability; 3) environmental complexity heightens demand for flexibility. Second, the strong path dependence we observe highlights habit and trust accumulation as central dynamics in hybrid human-robot service adoption, implying that short-run menu design and early experiences can have long-run effects on delivery type shares. Third, by analyzing choices in a field setting where users actively select the service type (rather than responding to hypothetical prompts), we provide practical evidence connecting lab-level psychological mechanisms to real-world behavior.

\subparagraph{Managerial implications.}
For platforms and operators designing hybrid human-robot systems, the results point to three actionable principles. 1) Context-aware channel design: promote robot delivery prominently for privacy-sensitive categories to strengthen perceptions of security and non-judgment. 2) Dynamic resource allocation and pricing: anticipate shifts toward human couriers during adverse weather and consider dynamic staffing or surge pricing to preserve service quality; pay attention to heterogeneous price sensitivities across delivery types and stations, which supports targeted price experiments. 3) Experience management to accelerate desirable adoption: because prior choices strongly predict subsequent choices, targeted trial incentives and immediate post-delivery promotions can accelerate habit formation for the corresponding delivery type.

\subparagraph{Limitations and future research.}
This study is subject to several limitations that also open avenues for future work. First, the evidence comes from campus Cainiao Stations, where demographics and operational structures may differ from broader urban or residential contexts, suggesting the need for replication in diverse settings and with alternative robot designs. Although we implement alternative specifications, the observational design cannot fully eliminate concerns about omitted variables. Experimental approaches could more directly test the underlying mechanisms. Finally, future research could extend our analysis to capture long-run dynamics, such as how habit formation influences market shares, cost structures, and welfare outcomes as the automation diffuses.

Taken together, our results indicate that consumer choice in hybrid human-robot delivery systems is systematic, psychologically grounded, and path-dependent. Designing service channels that are sensitive to privacy, value, and environmental context, and that proactively shape early user experiences, can improve both user experience and operational performance as automation continues to scale.

\newpage
\printbibliography

\appendix
\newpage
\section*{Appendix A: Additional Information} \label{sec:appendix A}
\setcounter{table}{0}
\setcounter{figure}{0}
\renewcommand{\thefigure}{A\arabic{figure}}  

\begin{figure}[H]
    \centering
    \includegraphics[width=0.9\linewidth]{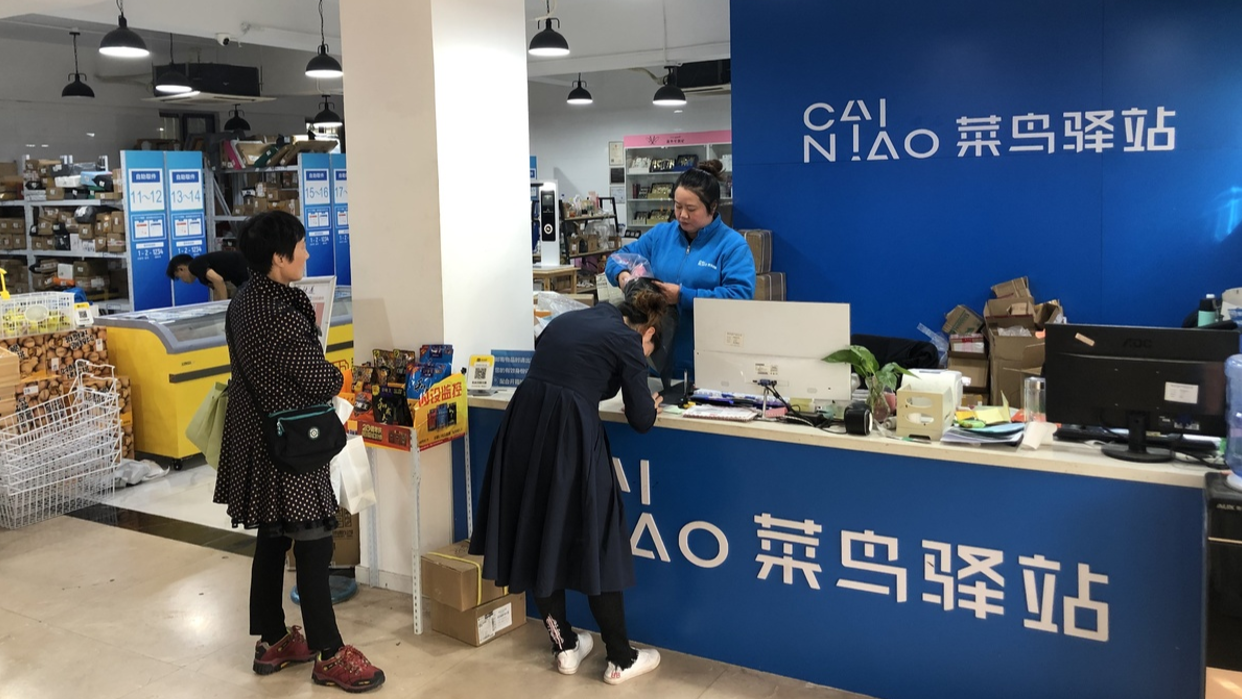}
    \caption{Cainiao Station}
    \label{fig:station}
\end{figure}

\begin{figure}[H]
    \centering
    \includegraphics[width=0.9\linewidth]{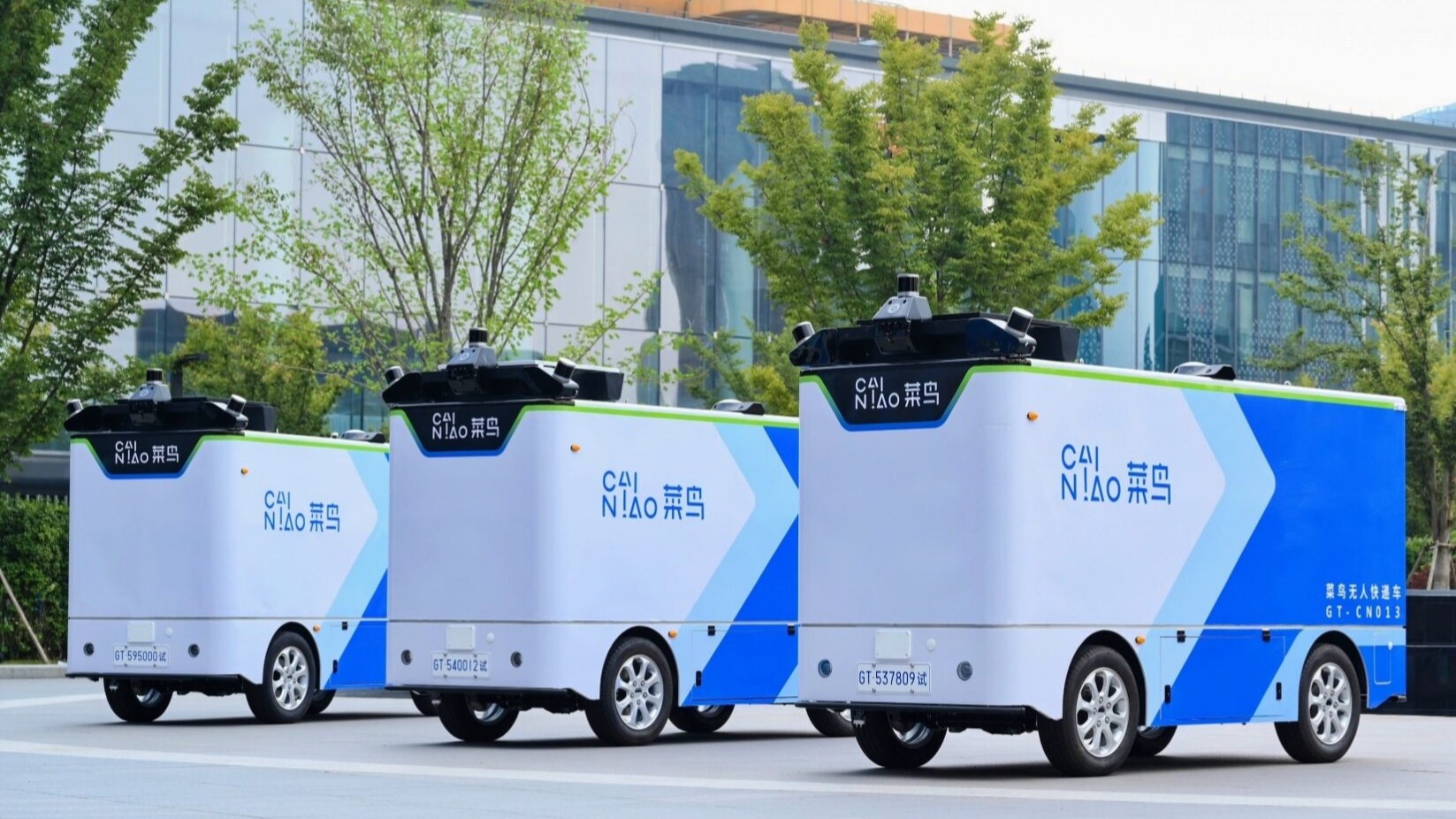}
    \caption{Cainiao Autonomous Delivery Robot}
    \label{fig:robot}
\end{figure}

\newpage
\section*{Appendix B: Tables for Robustness Checks} \label{sec:appendix B}
\setcounter{table}{0}
\setcounter{figure}{0}
\renewcommand{\thetable}{B\arabic{table}}  
\begin{table}[H]
\centering
\caption{Alternative models (Logit \& Probit)}

\footnotesize

\begin{tabular}{rcc}
\hline
\multicolumn{1}{l}{\multirow{2}{*}{}} & \multicolumn{2}{c}{\multirow{2}{*}{Dependent Variable: $delivery\_type$}}                                                     \\
\multicolumn{1}{l}{}                  & \multicolumn{2}{c}{}                                                                                                          \\ \cline{2-3} 
                                      & \begin{tabular}[c]{@{}c@{}}(1)\\ Logit\end{tabular}           & \begin{tabular}[c]{@{}c@{}}(2)\\ Probit\end{tabular}          \\ \hline
$const$                               & \begin{tabular}[c]{@{}c@{}}2.7700***\\ (0.0263)\end{tabular}  & \begin{tabular}[c]{@{}c@{}}2.7700***\\ (0.0263)\end{tabular}  \\
$is\_package\_sensitive$ & \begin{tabular}[c]{@{}c@{}}0.5992***\\ (0.0783)\end{tabular} & \begin{tabular}[c]{@{}c@{}}0.5992***\\ (0.0783)\end{tabular} \\
$\lg(package\_value)$                 & \begin{tabular}[c]{@{}c@{}}0.0649***\\ (0.0036)\end{tabular}  & \begin{tabular}[c]{@{}c@{}}0.0649***\\ (0.0036)\end{tabular}  \\
$is\_bad\_weather$                    & \begin{tabular}[c]{@{}c@{}}-0.1314***\\ (0.0097)\end{tabular} & \begin{tabular}[c]{@{}c@{}}-0.1314***\\ (0.0097)\end{tabular} \\
$lead\_time$                          & \begin{tabular}[c]{@{}c@{}}-0.1282***\\ (0.0014)\end{tabular} & \begin{tabular}[c]{@{}c@{}}-0.1282***\\ (0.0014)\end{tabular} \\
$lead\_time$                          & \begin{tabular}[c]{@{}c@{}}-0.0051***\\ (0.0007)\end{tabular} & \begin{tabular}[c]{@{}c@{}}-0.0051***\\ (0.0007)\end{tabular} \\
$delivery\_price$                     & \begin{tabular}[c]{@{}c@{}}-1.1686***\\ (0.0082)\end{tabular} & \begin{tabular}[c]{@{}c@{}}-1.1686***\\ (0.0082)\end{tabular} \\
$is\_female$                          & \begin{tabular}[c]{@{}c@{}}-0.3071***\\ (0.0100)\end{tabular} & \begin{tabular}[c]{@{}c@{}}-0.3071***\\ (0.0100)\end{tabular} \\
$is\_ios$                             & \begin{tabular}[c]{@{}c@{}}0.0585***\\ (0.0095)\end{tabular}  & \begin{tabular}[c]{@{}c@{}}0.0585***\\ (0.0095)\end{tabular}  \\
$is\_time\_sensitive$                 & \begin{tabular}[c]{@{}c@{}}0.0233*\\ (0.0125)\end{tabular}    & \begin{tabular}[c]{@{}c@{}}0.0233*\\ (0.0125)\end{tabular}    \\
$is\_South$                           & \begin{tabular}[c]{@{}c@{}}0.1163***\\ (0.0108)\end{tabular}  & \begin{tabular}[c]{@{}c@{}}0.1163***\\ (0.0108)\end{tabular}  \\
$is\_school\_South$                   & \begin{tabular}[c]{@{}c@{}}-0.5664***\\ (0.0115)\end{tabular} & \begin{tabular}[c]{@{}c@{}}-0.5664***\\ (0.0115)\end{tabular} \\
$is\_211$                             & \begin{tabular}[c]{@{}c@{}}1.0690***\\ (0.0167)\end{tabular}  & \begin{tabular}[c]{@{}c@{}}1.0690***\\ (0.0167)\end{tabular}  \\
\textbf{$is\_Monday$}                 & \begin{tabular}[c]{@{}c@{}}0.0150\\ (0.0174)\end{tabular}     & \begin{tabular}[c]{@{}c@{}}0.0150\\ (0.0174)\end{tabular}     \\
\textbf{$is\_Tuesday$}                & \begin{tabular}[c]{@{}c@{}}-0.0294*\\ (0.0176)\end{tabular}   & \begin{tabular}[c]{@{}c@{}}-0.0294*\\ (0.0176)\end{tabular}   \\
\textbf{$is\_Thursday$}               & \begin{tabular}[c]{@{}c@{}}0.0221\\ (0.0174)\end{tabular}     & \begin{tabular}[c]{@{}c@{}}0.0221\\ (0.0174)\end{tabular}     \\
\textbf{$is\_Friday$}                 & \begin{tabular}[c]{@{}c@{}}0.0687***\\ (0.0175)\end{tabular}  & \begin{tabular}[c]{@{}c@{}}0.0687***\\ (0.0175)\end{tabular}  \\
\textbf{$is\_Saturday$}  & \begin{tabular}[c]{@{}c@{}}0.0836***\\ (0.0175)\end{tabular} & \begin{tabular}[c]{@{}c@{}}0.0836***\\ (0.0175)\end{tabular} \\
\textbf{$is\_Sunday$}                 & \begin{tabular}[c]{@{}c@{}}0.1204***\\ (0.0169)\end{tabular}  & \begin{tabular}[c]{@{}c@{}}0.1204***\\ (0.0169)\end{tabular}  \\
\multirow{2}{*}{$Pseudo\; R^2$}         & \multirow{2}{*}{0.1679}                                       & \multirow{2}{*}{0.1679}                                       \\
                                      &                                                               &                                                               \\
\multirow{2}{*}{Observations}         & \multirow{2}{*}{241517}                                       & \multirow{2}{*}{241517}                                       \\
                                      &                                                               &                                                               \\ \hline
\multicolumn{3}{l}{\multirow{2}{*}{\small\textit{Note.} *$p<$0.1, ** $p<$0.05, *** $p<$0.01. Standard errors are in parentheses.}}                                    \\
\multicolumn{3}{l}{}                                                                                                                                                 
\end{tabular}

\label{tab:logit and probit}
\end{table}
\begin{table}[H]
\centering
\caption{Main Result with Two-way Fixed Effect}
\footnotesize
\label{tab:two-way effect}
\begin{tabular}{rcc}
\hline
\multicolumn{1}{l}{\multirow{2}{*}{}} & \multicolumn{2}{c}{\multirow{2}{*}{Dependent Variable: $delivery\_type$}}                                                     \\
\multicolumn{1}{l}{}                  & \multicolumn{2}{c}{}                                                                                                          \\ \cline{2-3} 
                                      & \begin{tabular}[c]{@{}c@{}}(1)\\ One-way FE\end{tabular}      & \begin{tabular}[c]{@{}c@{}}(2)\\ Two-way FE\end{tabular}      \\ \hline
$const$                               & \begin{tabular}[c]{@{}c@{}}0.9069***\\ (0.0040)\end{tabular}  & \begin{tabular}[c]{@{}c@{}}0.4708**\\ (0.2315)\end{tabular}   \\
$is\_package\_sensitive$ & \begin{tabular}[c]{@{}c@{}}0.1149***\\ (0.0143)\end{tabular}  & \begin{tabular}[c]{@{}c@{}}0.0231**\\ (0.0100)\end{tabular}   \\
$\lg(package\_value)$    & \begin{tabular}[c]{@{}c@{}}0.0097***\\ (0.0007)\end{tabular}  & \begin{tabular}[c]{@{}c@{}}0.0032***\\ (0.0005)\end{tabular}  \\
$is\_bad\_weather$                    & \begin{tabular}[c]{@{}c@{}}-0.0163***\\ (0.0022)\end{tabular} & \begin{tabular}[c]{@{}c@{}}-0.0014\\ (0.0016)\end{tabular}    \\
$lead\_time$                          & \begin{tabular}[c]{@{}c@{}}-0.0235***\\ (0.0002)\end{tabular} & \begin{tabular}[c]{@{}c@{}}-0.0065***\\ (0.0002)\end{tabular} \\
$lead\_time$                          & \begin{tabular}[c]{@{}c@{}}0.0038***\\ (0.0002)\end{tabular}  & \begin{tabular}[c]{@{}c@{}}-0.0003**\\ (0.0001)\end{tabular}  \\
$delivery\_price$        & \begin{tabular}[c]{@{}c@{}}-0.1489***\\ (0.0010)\end{tabular} & \begin{tabular}[c]{@{}c@{}}-0.0988***\\ (0.0009)\end{tabular} \\
$is\_Female$                          & \begin{tabular}[c]{@{}c@{}}-0.0590***\\ (0.0019)\end{tabular} &                                                               \\
$is\_ios$                             & \begin{tabular}[c]{@{}c@{}}0.0047**\\ (0.0018)\end{tabular}   &                                                               \\
$is\_time\_sensitive$                 & \begin{tabular}[c]{@{}c@{}}0.0020\\ (0.0024)\end{tabular}     &                                                               \\
$is\_South$                           & \begin{tabular}[c]{@{}c@{}}0.0178***\\ (0.0021)\end{tabular}  &                                                               \\
$is\_school\_South$                   & \begin{tabular}[c]{@{}c@{}}-0.1140***\\ (0.0022)\end{tabular} & \begin{tabular}[c]{@{}c@{}}0.1379\\ (0.3361)\end{tabular}     \\
$is\_211$                             & \begin{tabular}[c]{@{}c@{}}0.2101***\\ (0.0026)\end{tabular}  & \begin{tabular}[c]{@{}c@{}}1.0659***\\ (0.2059)\end{tabular}  \\
\multirow{2}{*}{$UserFE$}             & \multirow{2}{*}{No}                                           & \multirow{2}{*}{Yes}                                          \\
                                      &                                                               &                                                               \\
\multirow{2}{*}{$TimeFE$}             & \multirow{2}{*}{Yes}                                          & \multirow{2}{*}{Yes}                                          \\
                                      &                                                               &                                                               \\
\multirow{2}{*}{$R^2$}                & \multirow{2}{*}{0.1732}                                       & \multirow{2}{*}{0.0776}                                       \\
                                      &                                                               &                                                               \\
\multirow{2}{*}{Observations}         & \multirow{2}{*}{241517}                                       & \multirow{2}{*}{241517}                                       \\
                                      &                                                               &                                                               \\ \hline
\multicolumn{3}{l}{\multirow{2}{*}{\small\textit{Note.} *$p<$0.1, ** $p<$0.05, *** $p<$0.01. Standard errors are in parentheses.}}                                    \\
\multicolumn{3}{l}{}                                                                                                                                                 
\end{tabular}
\end{table}

\end{document}